\documentclass{article}
\usepackage{spconf,amsmath,graphicx,booktabs,amssymb}

\title{MR-SVS: Singing Voice Synthesis with Multi-Reference Encoder}
\name{Shoutong Wang$^{\star}$, Jinglin Liu$^{\star}$, Yi Ren$^{\star}$, Zhen Wang$^{\dagger}$, Changliang Xu$^{\dagger}$, Zhou Zhao$^{\star}$}
\address{$^{\star}$Zhejiang University\\
  $^{\dagger}$State Key Laboratory of Media Convergence Production Technology and Systems}
\begin{document}
%
\maketitle
\begin{abstract}
  Multi-speaker singing voice synthesis is to generate the singing voice sung by different speakers. To generalize to new speakers, previous zero-shot singing adaptation methods obtain the timbre of the target speaker with a fixed-size embedding from single reference audio. However, they face several challenges: 1) the fixed-size speaker embedding is not powerful enough to capture full details of the target timbre; 2) single reference audio does not contain sufficient timbre information of the target speaker; 3) the pitch inconsistency between different speakers also leads to a degradation in the generated voice. In this paper, we propose a new model called MR-SVS to tackle these problems. Specifically, we employ both a multi-reference encoder and a fixed-size encoder to encode the timbre of the target speaker from multiple reference audios. The Multi-reference encoder can capture more details and variations of the target timbre. Besides, we propose a well-designed pitch shift method to address the pitch inconsistency problem. Experiments indicate that our method outperforms the baseline method both in naturalness and similarity.  \footnote{Audio samples are posted at https://mr-svs.github.io}
\end{abstract}
\begin{keywords}
singing conversion, singing voice synthesis
\end{keywords}
\section{Introduction}
\label{sec:intro}

Singing is one of the predominant components of music production. Generating the singing voice of a target speaker by machine can highly reduce the cost of music production, while potentially creating new forms of music art. Training a separate singing model \cite{gu_bytesing_2020,blaauw_sequence--sequence_2020,lu_xiaoicesing_2020}, a voice conversion system \cite{deng_pitchnet_2020, chen_singing_2019, li_ppg-based_2020, nachmani_unsupervised_2019}, or fine-tuning an existing model \cite{blaauw_data_2019} can all reach the goal with good performance, but they either require a fair amount of well-labeled high-quality singing data of target speaker, additional training steps or occupy a large amount of storage for songs to synthesize. These properties limit their practical usage.

Zero-shot style singing adaptation methods \cite{ zhang_durian-sc_2020} which originate from \cite{jia_transfer_2019} are proposed to alleviate the problems introduced before. The main idea of zero-shot methods is to train a speaker encoder based on speaker verification methods, such as \cite{snyder_x-vectors_2018, wan_generalized_2019}. The pre-trained speaker encoder produces a speaker embedding with respect to the given reference audio. The speaker embedding is used to represent the target speaker timbre. Zero-shot style methods are fast and convenient because they do not need additional training steps or well-labeled high-quality data. These methods are suitable for usage in data-limited real-world conditions, but their effect is not as good as that of the fine-tuning methods \cite{chen_sample_2019}. Because  1) zero-shot methods can only use single reference audio which may not contain sufficient information of target timbre in different circumstances; 2) a single fixed-size embedding contains limited information and does not suffice to capture how the target speaker performs various singing techniques. 

Inspired by recent speech adaptation methods \cite{choi_attentron_2020, chen_adaspeech_2020} at different granularities, in this paper, we introduce the multi-head attention-based multi-reference encoder to the zero-shot multi-speaker singing voice synthesis system. Namely, the multi-reference encoder based singing voice synthesis (MR-SVS) system. Besides, we also propose a pitch shift method. Specifically, 1) our multi-reference encoder computes an exclusive embedding for every generated singing frame and complements the details missed by the overall fixed-size embedding. 2) Our multi-reference encoder can exploit multiple reference audios together to generate more accurate and precise embeddings. 3) The multi-head attention mechanism helps our multi-reference encoder to attend to different properties of singing voice production from all the reference audios. 4) Our pitch shift method shifts the pitch to conform to the average pitch of the target speaker to address the pitch inconsistency problem which is especially severe with different genders.

We conduct experiments on our proprietary singing corpus with different out-of-set reference speech audios. We measure and compare the performance of models using Mean Opinion Scores (MOS) from a crowd of testers. The evaluation results by comparing the naturalness and similarity show that our proposed model outperforms the baseline model\cite{zhang_durian-sc_2020}. The ablation studies address the effectiveness of our proposed model components and methods. Besides keeping the convenience and efficiency of zero-shot methods, our method is also highly controllable, which may lead to new ways of music production.

\section{MR-SVS}
\label{sec:mr-svc}

\subsection{Overall Acoustic Model}

\begin{figure}[t]
  \centering
  \includegraphics[width=\hsize]{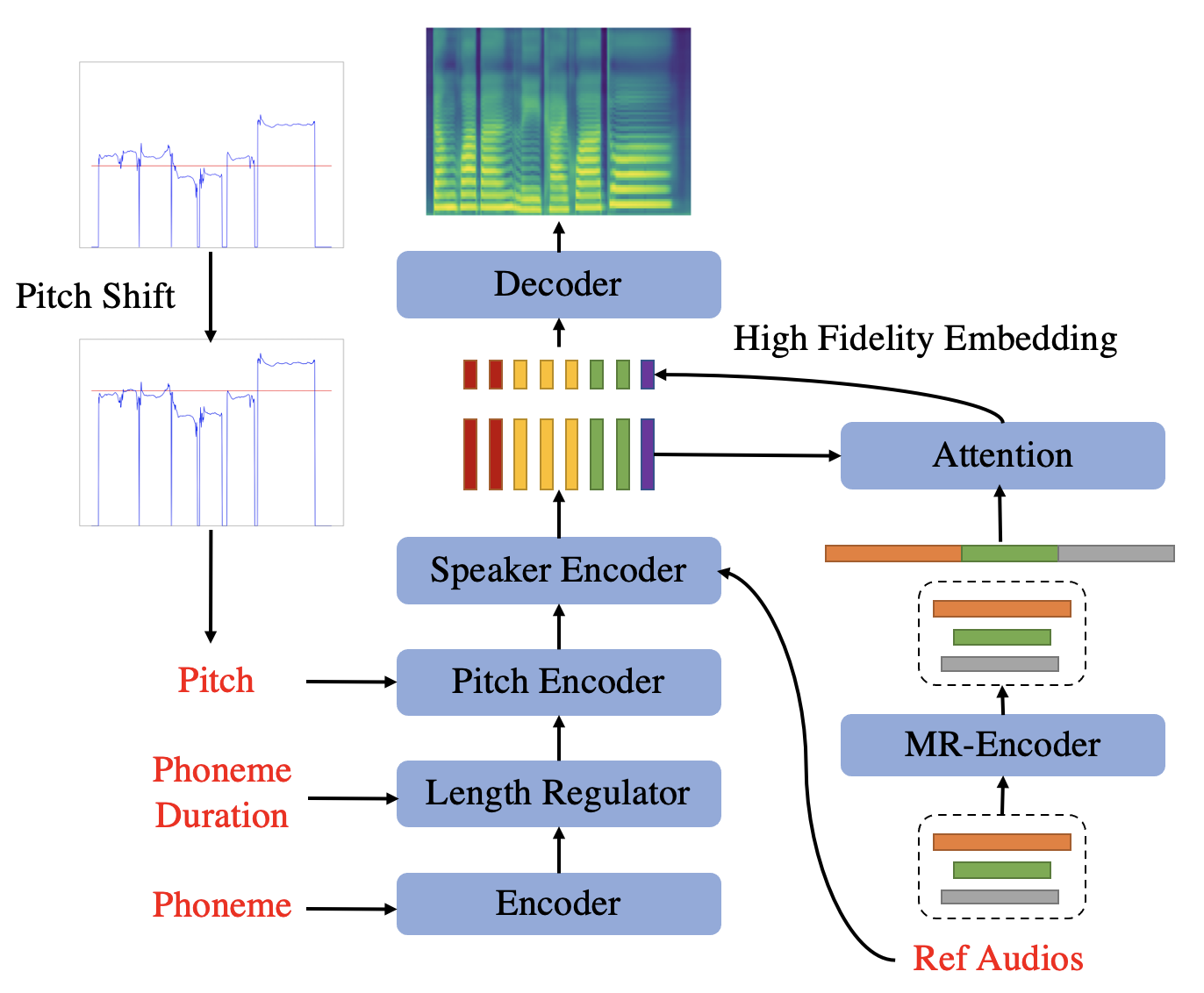}
  \caption{Overall synthesis process. Pitch shift is only performed during inference. Model inputs are colored in red.}
  \label{fig:overall_architecture}
\end{figure}

Singing synthesis differs from speech synthesis since singing synthesis is constrained by the music score while the prosody of speech synthesis requires no particular constraints. As a substitution to the music score, we provide phoneme, phoneme duration, and pitch contour as input to our singing synthesis system.

We modify the FastSpeech 2\cite{ren_fastspeech_2021} model to meet the requirements for singing synthesis. Our modified model includes (1) an encoder that encodes phoneme information, (2) a length regulator that expands encoder outputs according to the phoneme duration to generate the frame-level hidden states, (3) a pitch encoder to encode frame-level pitch information to the hidden states, and (4) a decoder to generate target mel-spectrogram frames from the hidden states. It takes phoneme, phoneme duration, and pitch as input, along with the embeddings from fixed-size speaker encoder and multi-reference encoder, generates mel-spectrogram as input to the subsequent vocoder. Compared with \cite{ren_fastspeech_2021}, we omit the variance adaptor since its functionality is completely replaced by the input information. Our model is trained to minimize the reconstruction loss of the mel-spectrogram. The overall synthesis process is illustrated in Figure~\ref{fig:overall_architecture}.

\subsection{Speaker Encoders}

\subsubsection{Multi-Head Multi-Reference Embedding Encoder}

As shown in Figure~\ref{fig:overall_architecture}, the inputs to multi-head multi-reference embedding encoder are several random reference audio files from the target speaker. We first convert them to the mel-spectrograms. Converted reference spectrograms are passed through two convolutional layers followed by two bidirectional LSTM layers to accompany context information. We flatten the outputs of every reference audio to get a long sequence $S$. Given the expanded hidden state $h_i$, we compute the attention query $Q_i = h_i W_q \in \mathbb{R}^{d_m}$, key $K = S W_k \in \mathbb{R}^{L_r \times d_m}$, and value $V = S W_v \in \mathbb{R}^{L_r \times d_m}$. $L_r$ is the total length of flattened LSTM outputs, $d_m$ is the length of embedding, and $W$s are trainable mapping matrices. The hidden states contain phoneme, context, and pitch information.  We calculate the high-fidelity embedding $e_i$ for each hidden state $h_i$ as:
\begin{equation}
  e_i = softmax(\frac{Q_i K^T}{\sqrt{d_m}}) V.
\end{equation}

To take different aspects of singing voice production into consideration, it is natural for us to adopt the multi-head attention \cite{vaswani_attention_2017} mechanism. Ideally, it can attend to different properties on different attention heads, which will help to generate better auxiliary embeddings.

This multi-reference encoder (MR-Encoder) is trained along with the singing synthesis system. For each target singing audio sample, we randomly choose several audios from the same speaker and regard them as the reference audio to produce the high fidelity embeddings for each singing frame.

\subsubsection{Fixed-Size Speaker Encoder}

The fixed-size speaker encoder is used to generate a fixed-size speaker representation from the target reference audio. We follow \cite{wan_generalized_2019}, whose original purpose is for speaker verification. The network proposed in \cite{wan_generalized_2019} is trained to optimize a generalized end-to-end speaker verification loss, so that embeddings of utterances from the same speaker have high cosine similarity, while those utterances from different speakers are far apart in embedding space. We adopt the pre-trained model in package Resemblyzer\footnote{https://github.com/resemble-ai/Resemblyzer}. The parameters will be frozen all the time. During training, the target audio is fed into this encoder. During inference, reference audios are fed into the encoder and the results are averaged to represent the target speaker.

\subsection{Pitch Shift}

It's very common to see a large gap between the source and target speaker's pitch range, especially when performing singing conversion with source pitch and target speaker from different genders. During training, the ground truth pitch is provided to generate the mel-spectrogram of the training utterances. During inference, the ground truth pitch may not work well since the pitch gap will lead to significant deviation. To alleviate this problem, we compute the average $f0$ of reference audios over all voiced frames, regarding it as the scalar reference $f0$ of the target speaker. We shift the source $f0$ so that the average $f0$ consumed by the pitch encoder is the same as the scalar reference $f0$.
\begin{equation}
  {f0}_{input} = {f0}_{s} + mean({f0}_{t}) - mean({f0}_{s})
\end{equation}
${f0}_{t}$ is the pitch sequence of target reference audios. ${f0}_{s}$ is the pitch sequence of the source audio. The calculation result sequence ${f0}_{input}$ will be the pitch fed into the pitch encoder during inference. We have also set a lower bound of ${f0}_{input}$ in the case when the pitch is too low to be modeled by our pitch encoder.

\subsection{Modified Vocoder}

Originally proposed Parallel WaveGAN \cite{yamamoto_parallel_2020} vocoder takes mel-spectrogram as conditional to produce speech waveforms. To model the pitch variations better, we take both pitch contours and mel-spectrogram as vocoder input. We discretize and represent each frame-level pitch with an 80-dimensional pitch embedding. We concatenate pitch embedding and mel-spectrogram together and pass them through a Fully-Connected layer before the upsampling procedure.

\section{Experiments}

\subsection{Experimental Settings}

\subsubsection{Dataset}

Our proprietary singing corpus contains 1810 Mandarin pop songs split into 69845 singing segments. The corpus is recorded without any accompaniment and contains only human singing voice. There are 93 singers in total among whom 56 are female singers and 37 are male singers. The total length of all songs is about 82.4 hours. From the singing dataset 4 songs (120 clips) are chosen for testing. The test songs are sung by 2 female and 2 male singers have no overlap with training and validation set. Among singing segments left, 400 clips are used for validation and else for training. We transform the raw waveform into 80-dimensional mel-spectrograms and set frame size and hop size to 512 and 128 with respect to the sample rate 22050Hz.

\subsubsection{Data Preprocess}

 We extract the phoneme duration with MFA \cite{mcauliffe_montreal_2017}, an open-source system for speech-text alignment with good performance, which can be trained on paired text-audio corpus without any manual alignment annotations. We convert the alignment results generated by MFA \cite{mcauliffe_montreal_2017} to phoneme durations. We also extract $f0$ from the raw waveform by World vocoder \cite{morise_world_2016} with the same hop size of target mel-spectrograms to obtain the pitch of each frame. 

\subsubsection{Model Parameters}



The acoustic model consists of 4 feed-forward Transformer\cite{ren_fastspeech_2019} blocks both for the phoneme encoder and mel-spectrogram decoder. The hidden dimension of the phoneme embedding, speaker embedding, pitch embedding, the hidden in self-attention and others is 256. The number of attention heads, the feed-forward filter size and kernel size are set to 2, 1024 and 9 respectively.

The multi-reference encoder consists of 2 convolutional layers with filter size and kernel size of 512 and 3 respectively, and followed by a 2-layer bidirectional LSTM layer with 256 cells. The hidden dimension (dimension of Q, K, V and high-fidelity embedding) in multi-head attention and the number of attention heads are 256 and 8. The fixed-size speaker encoder follows the architecture in Resemblyzer.

We train our model on an NVIDIA RTX 2080Ti with dynamic batch size. We use the AdamW optimizer \cite{loshchilov_2018_decoupled} with $\beta_1 = 0.9, \beta_2 = 0.98, \epsilon = 10^{-9}$ and follow the same learning rate schedule in \cite{vaswani_attention_2017}. It takes 2M steps for training to complete. We train our vocoder on our singing corpus for 1M steps.

\subsubsection{Evaluation Methods}

For evaluation, Mean Opinion Scores (MOS) on naturalness and similarity to the target speaker are evaluated. The scale of MOS is set between 1 to 5 with 5 representing the best performance and 1 the worst. 11 evaluators participate in our listening test. In tables we present MOS with a 95\% confidence interval. 2 male and 2 female speakers in AISHELL 3 \cite{shi_aishell-3_2020} dataset are chosen as target speakers. We use 8 reference audio clips for each speaker. Our 4 test songs (120 clips) are sung by 2 female and male singers respectively. We can evaluate inter-gender and intra-gender singing voice conversion with this setting.

\subsection{Results}

\begin{table}[th]
  \caption{Evaluation results. The naturalness scores are evaluated by comparing with the vocoder-generated GTs.}
  \label{tab:evaluation_result}
  \centering
  \begin{tabular}{ c c c }
    \toprule
    \textbf{Model} & 
    \textbf{Naturalness} &
    \textbf{Similarity}\\
    \midrule
    \textbf{\emph{Baseline}} \cite{zhang_durian-sc_2020}           & $4.23 \pm 0.02 $  & $3.91 \pm 0.02$             \\
    \textbf{\emph{Multi-ref-only}}   & $4.27 \pm 0.01$  & $3.73 \pm 0.02$               \\
    \textbf{\emph{Single-head}}          & $4.30 \pm 0.01$  & $3.97 \pm 0.02$       \\
    \textbf{\emph{MR-SVS}}       & $4.42 \pm 0.01$  & $4.05 \pm 0.01$              \\
    \bottomrule
  \end{tabular}
  
\end{table}

\subsubsection{Performance Evaluation}

Our \textbf{\emph{baseline}} model uses the same conversion method as \cite{zhang_durian-sc_2020}. This model is trained with only the fixed-size speaker encoder, while our \textbf{\emph{MR-SVS}} model has an additional 8-head multi-reference encoder. The fixed-size embedding is the average embedding of reference audios for both of them. The MOS of naturalness and similarity listed in Table~\ref{tab:evaluation_result} denote that the \textbf{\emph{MR-SVS}} model improves over the \textbf{\emph{baseline}} model both in similarity and naturalness. Constructing each singing frame by exploiting reference audios at a high granularity shows its advantage over the conventional single fixed-size embedding model.

\subsubsection{Ablation Studies}

To illustrate the effectiveness of components in our model, besides the \textbf{\emph{baseline}} and \textbf{\emph{MR-SVS}}, we train two additional models: 1) \textbf{\emph{multi-ref-only}} model is trained with only the multi-reference encoder; 2) \textbf{\emph{single-head}} model is trained with both a fixed-size speaker encoder and a single-head multi-reference encoder.

The MOS of additional models are also listed in Table~\ref{tab:evaluation_result}. Our \textbf{\emph{MR-SVS}} model outperforms all other models both in similarity and naturalness.

The \textbf{\emph{baseline}} model and the \textbf{\emph{multi-ref-only}} model achieve better scores in similarity and naturalness respectively, addressing the importance of fixed-size speaker encoder for the overall target timbre and multi-reference encoder for the subtle singing details. The Multi-reference-encoder-only method does not suffice to capture the target timbre well. Though not intended originally, the contrastive learning method in \cite{wan_generalized_2019} can learn a good representation for the target timbre.

While the \textbf{\emph{single-head}} model already achieves a better score in similarity and naturalness over the \textbf{\emph{baseline}} model and the \textbf{\emph{multi-ref-only}} model, showing the power of using both encoders together, the multi-head attention mechanism can still lead to an uplift in both measures over the \textbf{\emph{single-head}} model. The flexibility of multi-head attention can help with singing synthesis as expected.

\subsubsection{Inter-Gender and Intra-Gender Conversion}

\begin{table}[th]
  \caption{Inter and intra-gender evaluation results, M and F represents male and female respectively.}
  \label{tab:inter_and_intra_gender_conversion}
  \centering
  \begin{tabular}{ c c c c }
    \toprule
    \textbf{Source} & 
    \textbf{Target} &
    \textbf{Naturalness} &
    \textbf{Similarity}\\
    \midrule
    \multicolumn{4}{c}{\textbf{With Pitch Shift}} \\
    \midrule
    M & M & $4.42 \pm 0.03 $  & $4.18 \pm 0.03 $             \\
    M & F & $4.40 \pm 0.03 $  & $3.70 \pm 0.05 $               \\
    F & M & $4.36 \pm 0.02 $  & $4.08 \pm 0.03 $       \\
    F & F & $4.51 \pm 0.02 $  & $4.24 \pm 0.03 $              \\
    \midrule
    \multicolumn{4}{c}{\textbf{Without Pitch Shift}} \\
    \midrule
    M & M & $4.25 \pm 0.04 $  & $4.02 \pm 0.04 $             \\
    M & F & $4.19 \pm 0.04 $  & $3.51 \pm 0.07 $             \\
    F & M & $4.20 \pm 0.03 $  & $3.93 \pm 0.04 $             \\
    F & F & $4.27 \pm 0.04 $  & $4.06 \pm 0.03 $             \\
    \bottomrule
  \end{tabular}
\end{table}

We address the importance of the pitch shift method here. The MOS of inter-gender and intra-gender conversion are listed in Table~\ref{tab:inter_and_intra_gender_conversion}. These samples are all generated by our \textbf{\emph{MR-SVS}} model. The scores show that with pitch shift, our model accomplishes a significant evaluation uplift compared to those without it both in naturalness and similarity, even for intra-gender conversion. The results imply that though not as severe as inter-gender conversion, the pitch inconsistency problem also exists and leads to conversion degradation for speakers of the same gender.

Among our music-score-like inputs, apparently, the phoneme and phoneme duration do not contain timbre-related information. Since the conversion performance between different genders is still unable to match intra-gender ones, we can conclude from this phenomenon that even the shifted pitch contours may contain subtle singing techniques specific to different genders and these techniques cannot be well performed by the target timbre.

\subsubsection{Controllable Singing Voice Synthesis}

\begin{figure}[t]
  \centering
  \includegraphics[width=\hsize]{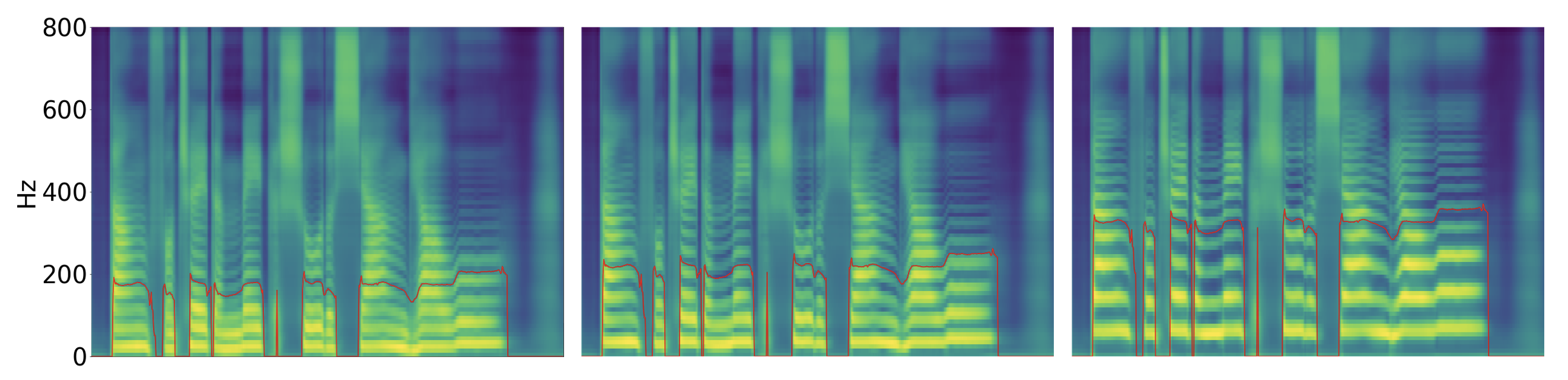}
  \caption{Input pitch of spectrograms increases from left to right with same phoneme, phoneme duration and speaker embedding settings. The pitch contours are drawn in red.}
  \label{fig:control_singing}
\end{figure}

We can change the melody or content of our synthesized singing segments by perturbing the corresponding inputs. The mel-spectrograms with overall modification on pitch input are shown in Figure~\ref{fig:control_singing}. In addition to the overall control, we can also change the phoneme, phoneme duration, and pitch input at desired proportion. This feature may lead to new forms of music production. 

\section{Conclusions}

In this paper, we proposed a zero-shot style multi-speaker singing voice synthesis system called MR-SVS which is suitable for usage in low-resource circumstances. Besides the conventional fixed-size speaker encoder, we introduced a new multi-head attention-based multi-reference encoder to generate high-fidelity embedding for each singing frame. We also highly alleviated the pitch inconsistency problem by adopting the pitch shift method. Experiments show that our proposed model can generate high-quality singing voices for the target speaker with a great uplift in naturalness and similarity while retaining the convenience and speed of conventional zero-shot methods. The generation process can also be manually controlled.

\vfill\pagebreak



\bibliographystyle{IEEEbib}
\bibliography{t}

\begin{thebibliography}{10}

\bibitem{gu_bytesing_2020}
Yu~Gu, Xiang Yin, Yonghui Rao, Yuan Wan, Benlai Tang, Yang Zhang, Jitong Chen,
  Yuxuan Wang, and Zejun Ma,
\newblock ``Bytesing: {A} chinese singing voice synthesis system using duration
  allocated encoder-decoder acoustic models and wavernn vocoders,''
\newblock in {\em {ISCSLP} 2021}. 2021, pp. 1--5, {IEEE}.

\bibitem{blaauw_sequence--sequence_2020}
Merlijn Blaauw and Jordi Bonada,
\newblock ``Sequence-to-{Sequence} {Singing} {Synthesis} {Using} the
  {Feed}-{Forward} {Transformer},''
\newblock in {\em {ICASSP} 2020}, May 2020, pp. 7229--7233,
\newblock ISSN: 2379-190X.

\bibitem{lu_xiaoicesing_2020}
Peiling Lu, Jie Wu, Jian Luan, Xu~Tan, and Li~Zhou,
\newblock ``Xiaoicesing: {A} high-quality and integrated singing voice
  synthesis system,''
\newblock in {\em Interspeech 2020}. 2020, pp. 1306--1310, {ISCA}.

\bibitem{deng_pitchnet_2020}
Chengqi Deng, Chengzhu Yu, Heng Lu, Chao Weng, and Dong Yu,
\newblock ``Pitchnet: Unsupervised singing voice conversion with pitch
  adversarial network,''
\newblock in {\em ICASSP 2020}, 2020, pp. 7749--7753.

\bibitem{chen_singing_2019}
Xin Chen, Wei Chu, Jinxi Guo, and Ning Xu,
\newblock ``Singing voice conversion with non-parallel data,''
\newblock in {\em {MIPR} 2019}. 2019, pp. 292--296, {IEEE}.

\bibitem{li_ppg-based_2020}
Zhonghao Li, Benlai Tang, Xiang Yin, Yuan Wan, Ling Xu, Chen Shen, and Zejun
  Ma,
\newblock ``{PPG}-based singing voice conversion with adversarial
  representation learning,''
\newblock {\em arXiv:2010.14804 [cs, eess]}, Oct. 2020,
\newblock arXiv: 2010.14804.

\bibitem{nachmani_unsupervised_2019}
Eliya Nachmani and Lior Wolf,
\newblock ``Unsupervised singing voice conversion,''
\newblock in {\em Interspeech 2019}. 2019, pp. 2583--2587, {ISCA}.

\bibitem{blaauw_data_2019}
Merlijn Blaauw, Jordi Bonada, and Ryunosuke Daido,
\newblock ``Data efficient voice cloning for neural singing synthesis,''
\newblock in {\em ICASSP 2019}, 2019, pp. 6840--6844.

\bibitem{zhang_durian-sc_2020}
Liqiang Zhang, Chengzhu Yu, Heng Lu, Chao Weng, Chunlei Zhang, Yusong Wu, Xiang
  Xie, Zijin Li, and Dong Yu,
\newblock ``Durian-sc: Duration informed attention network based singing voice
  conversion system,''
\newblock in {\em Interspeech 2020}. 2020, pp. 1231--1235, {ISCA}.

\bibitem{jia_transfer_2019}
Ye~Jia, Yu~Zhang, Ron Weiss, Quan Wang, Jonathan Shen, Fei Ren, zhifeng Chen,
  Patrick Nguyen, Ruoming Pang, Ignacio Lopez~Moreno, and Yonghui Wu,
\newblock ``Transfer learning from speaker verification to multispeaker
  text-to-speech synthesis,''
\newblock in {\em NeurIPS}, 2018, vol.~31.

\bibitem{snyder_x-vectors_2018}
David Snyder, Daniel Garcia-Romero, Gregory Sell, Daniel Povey, and Sanjeev
  Khudanpur,
\newblock ``X-{Vectors}: {Robust} {DNN} {Embeddings} for {Speaker}
  {Recognition},''
\newblock in {\em ICASSP 2018}. Apr. 2018, pp. 5329--5333, IEEE.

\bibitem{wan_generalized_2019}
Li~Wan, Quan Wang, Alan Papir, and Ignacio Lopez{-}Moreno,
\newblock ``Generalized end-to-end loss for speaker verification,''
\newblock in {\em {ICASSP} 2018}. 2018, pp. 4879--4883, {IEEE}.

\bibitem{chen_sample_2019}
Yutian Chen, Yannis~M. Assael, Brendan Shillingford, David Budden, Scott~E.
  Reed, Heiga Zen, Quan Wang, Luis~C. Cobo, Andrew Trask, Ben Laurie,
  {\c{C}}aglar G{\"{u}}l{\c{c}}ehre, A{\"{a}}ron van~den Oord, Oriol Vinyals,
  and Nando de~Freitas,
\newblock ``Sample efficient adaptive text-to-speech,''
\newblock in {\em {ICLR} 2019}, 2019.

\bibitem{choi_attentron_2020}
Seungwoo Choi, Seungju Han, Dongyoung Kim, and Sungjoo Ha,
\newblock ``Attentron: Few-shot text-to-speech utilizing attention-based
  variable-length embedding,''
\newblock in {\em Interspeech 2020}. 2020, pp. 2007--2011, {ISCA}.

\bibitem{chen_adaspeech_2020}
Mingjian Chen, Xu~Tan, Bohan Li, Yanqing Liu, Tao Qin, Sheng Zhao, and Tie-Yan
  Liu,
\newblock ``{AdaSpeech}: {Adaptive} {Text} to {Speech} for {Custom} {Voice},''
\newblock Sept. 2020.

\bibitem{ren_fastspeech_2021}
Yi~Ren, Chenxu Hu, Xu~Tan, Tao Qin, Sheng Zhao, Zhou Zhao, and Tie{-}Yan Liu,
\newblock ``Fastspeech 2: Fast and high-quality end-to-end text to speech,''
\newblock in {\em {ICLR} 2021}, 2021.

\bibitem{vaswani_attention_2017}
Ashish Vaswani, Noam Shazeer, Niki Parmar, Jakob Uszkoreit, Llion Jones,
  Aidan~N Gomez, \L~ukasz Kaiser, and Illia Polosukhin,
\newblock ``Attention is all you need,''
\newblock in {\em NeurIPS}, 2017, vol.~30.

\bibitem{yamamoto_parallel_2020}
Ryuichi Yamamoto, Eunwoo Song, and Jae{-}Min Kim,
\newblock ``Parallel wavegan: {A} fast waveform generation model based on
  generative adversarial networks with multi-resolution spectrogram,''
\newblock in {\em {ICASSP} 2020}. 2020, pp. 6199--6203, {IEEE}.

\bibitem{mcauliffe_montreal_2017}
Michael McAuliffe, Michaela Socolof, Sarah Mihuc, Michael Wagner, and Morgan
  Sonderegger,
\newblock ``Montreal forced aligner: Trainable text-speech alignment using
  kaldi,''
\newblock in {\em Proc. Interspeech 2017}, 2017, pp. 498--502.

\bibitem{morise_world_2016}
Masanori MORISE, Fumiya YOKOMORI, and Kenji OZAWA,
\newblock ``World: A vocoder-based high-quality speech synthesis system for
  real-time applications,''
\newblock {\em IEICE Transactions on Information and Systems}, vol. E99.D, no.
  7, pp. 1877--1884, 2016.

\bibitem{ren_fastspeech_2019}
Yi~Ren, Yangjun Ruan, Xu~Tan, Tao Qin, Sheng Zhao, Zhou Zhao, and Tie{-}Yan
  Liu,
\newblock ``Fastspeech: Fast, robust and controllable text to speech,''
\newblock in {\em NeurIPS}, 2019, pp. 3165--3174.

\bibitem{loshchilov_2018_decoupled}
Ilya Loshchilov and Frank Hutter,
\newblock ``Decoupled weight decay regularization,''
\newblock in {\em {ICLR} 2019}, 2019.

\bibitem{shi_aishell-3_2020}
Yao Shi, Hui Bu, Xin Xu, Shaoji Zhang, and Ming Li,
\newblock ``{AISHELL}-3: {A} {Multi}-speaker {Mandarin} {TTS} {Corpus} and the
  {Baselines},''
\newblock {\em arXiv:2010.11567 [cs, eess]}, Oct. 2020,
\newblock arXiv: 2010.11567.

\end{thebibliography}

\end{document}